%% file: main.tex
\documentclass[sigconf]{acmart}

\usepackage[utf8]{inputenc}

\input{definitions}
\input{packages}
\input{meta}
\begin{document}

\acmSubmissionID{1518}



\title[Divergence Meets Consensus: A Multi-Source Negative Sampling \\ Framework for Sequential Recommendation] 
{Divergence Meets Consensus: A Multi-Source Negative Sampling \\ Framework for Sequential Recommendation}


\settopmatter{authorsperrow=4}
\input{sections/authors}
\input{sections/abstract.tex}


\maketitle
\acresetall
\input{sections/introduction.tex}
\input{sections/P2.tex}
\input{sections/method.tex}

\input{sections/experiments.tex}

\input{sections/related_work.tex}

\input{sections/conclusion.tex}

\input{sections/acknowledgement.tex}
\clearpage
\bibliographystyle{ACM-Reference-Format}
\balance
\bibliography{main}

\appendix
\end{document}

%% file: definitions.tex
\AtBeginDocument{%
  \providecommand\BibTeX{{%
    \normalfont B\kern-0.5em{\scshape i\kern-0.25em b}\kern-0.8em\TeX}}}

\usepackage{xcolor} 
\usepackage{pifont}

\definecolor{myblue}{HTML}{cfeaf1}
\definecolor{myblue0}{HTML}{D7F9F0}
\definecolor{myred}{HTML}{f6cae5}

\newcommand{\model}{MDCNS}

\newcommand{\header}[1]{\vspace*{1mm}\noindent{\textbf{#1}}}

\newcommand{\yes}{\textbf{\textcolor{green!60!black}{\ding{51}}}}
\newcommand{\no}{\textbf{\textcolor{red}{\ding{55}}}}
\newcommand{\oca}{\textbf{\textcolor{blue}{\ding{109}}}}

%% file: packages.tex
\usepackage{amsmath}
\usepackage{amssymb}
\usepackage{balance}
\usepackage{amsfonts}
\usepackage{mathrsfs}
\usepackage{amsthm}
\usepackage{bbm}
\usepackage{array}
\usepackage{booktabs}
\usepackage{graphicx}
\usepackage{graphics}
\usepackage{epsfig}
\usepackage{xcolor}
\usepackage{colortbl}
\usepackage{multirow}
\usepackage{float}
\usepackage{balance}
\usepackage{acronym}
\usepackage{placeins}
\usepackage[inline]{enumitem}
\usepackage{fancyhdr}
\usepackage{threeparttable}
\usepackage{subfig}
\usepackage[ruled,linesnumbered]{algorithm2e}
\usepackage[noend]{algpseudocode}
\usepackage{adjustbox}
\usepackage{tikz}
\usepackage{pgfplots}
\pgfplotsset{compat=1.17}
\usepackage{caption}
\usepackage{listings}
\usepackage{tcolorbox}
\usepackage[title]{appendix}

\setitemize[1]{itemsep=0pt,partopsep=0pt,parsep=\parskip,topsep=5pt}

\usepackage[subfigure]{tocloft}

\usepackage{arydshln}
\usepackage{tabularx}

\usepackage{hyperref}
\usepackage[capitalise]{cleveref}

%% file: meta.tex
\copyrightyear{2026}
\acmYear{2026}
\setcopyright{cc}
\setcctype{by}
\acmConference[SIGIR '26]{Proceedings of the 49th International ACM SIGIR Conference on Research and Development in Information Retrieval}{July 20--24, 2026}{Melbourne, VIC, Australia}
\acmBooktitle{Proceedings of the 49th International ACM SIGIR Conference on Research and Development in Information Retrieval (SIGIR '26), July 20--24, 2026, Melbourne, VIC, Australia}
\acmDOI{10.1145/3805712.3809555}
\acmISBN{979-8-4007-2599-9/2026/07}

%% file: sections/authors.tex
\author{Yuanzi Li}
\affiliation{%
\institution{\mbox{Renmin University of China}}
  \city{Beijing}
  \country{China}
}
\email{liyuanzi0313@outlook.com}

\author{Lingjie Wang}
\affiliation{%
  \institution{Shandong University}
  \city{Qingdao}
  \country{China}
}
\email{lingjie.wang@mail.sdu.edu.cn}

\author{Jingyu Zhao}
\affiliation{%
  \institution{Shandong University}
  \city{Qingdao}
  \country{China}
}
\email{jingyu.zhao@mail.sdu.edu.cn}


\author{Zihang Tian}
\affiliation{%
\institution{\mbox{Renmin University of China}}
  \city{Beijing}
  \country{China}
}
\email{zihangtian@ruc.edu.cn}

\author{Yuhan Wang}
\affiliation{%
  \institution{Peking University}
  \city{Beijing}
  \country{China}
}
\email{fzzh040114@gmail.com}

\author{Lei Wang}
\affiliation{%
\institution{\mbox{Renmin University of China}}
  \city{Beijing}
  \country{China}
}
\email{wanglei154@ruc.edu.cn}

\author{Xu Chen}
\authornote{Corresponding author}
\affiliation{%
\institution{\mbox{Renmin University of China}}
\city{Beijing}
\country{China}
}
\email{xu.chen@ruc.edu.cn}

\renewcommand{\shortauthors}{Yuanzi Li et al.}

%% file: sections/abstract.tex
\begin{abstract}
Negative sampling is significant for training sequential recommendation models under implicit feedback. The predominant strategy, self-guided hard negative sampling, selects negatives based on the model’s current state but suffers from three limitations:
(1) the coupling between sampling and model updates triggers a vicious cycle that drives the model into local optima;
(2) relying on current model parameters narrows sampling to a small region of the item space, reducing diversity and harming generalization;
(3) identifying a hard negative requires scoring the entire candidate pool, causing substantial computational overhead with minimal information gain.

To address these challenges, we propose \textbf{MDCNS} (\textbf{\underline{M}}ulti-source \textbf{\underline{D}}ivergence-\textbf{\underline{C}}onsensus for \textbf{\underline{N}}egative \textbf{\underline{S}}ampling), a novel "Teacher-Peer-Self" framework inspired by Vygotsky's Zone of Proximal Development (ZPD) theory.
The proposed method comprises three components, including \emph{multi-source scoring}, \emph{divergence re-ranking}, and \emph{consensus distillation}.
Firstly, multi-source scoring incorporates peer and ensemble teacher models to inject external negative signals and break the self-reinforcement loop. Then, divergence re-ranking exploits prediction discrepancy between self and peer models to enhance sampling diversity. Finally, consensus distillation aligns the self model with the teacher via KL divergence, simultaneously improving computational cost utilization.
Extensive experiments on six real-world datasets and five backbone models show that \model\  consistently outperforms state-of-the-art negative sampling methods, demonstrating strong effectiveness and generalization.
Code is available at \url{https://github.com/Lyz103/SIGIR26-MDCNS}.
\end{abstract}

%% file: sections/introduction.tex
\vspace{-1em}
\section{Introduction}
\header{Sequential Recommendation.}
Recommender systems have become a cornerstone technology in modern information access, and are now deeply integrated into a wide range of application scenarios, such as e-commerce~\cite{hu2018reinforcement}, social media~\cite{yuan2019simple}, and other online services and platforms~\cite{qi2020privacy}.
Among the diverse recommendation paradigms, sequential recommendation (SR) has recently emerged as a central research topic, drawing sustained interest from both academic and industrial communities.
The fundamental goal of SR is to exploit users’ historical interaction sequences to infer and anticipate their upcoming preferences and potential future behaviors.
In most real-world systems, SR models are trained on implicit feedback (e.g., clicks and views), which makes it crucial to construct reliable sets of positive and negative instances for effective learning.
Because implicit feedback predominantly reflects positive user signals, negative sampling becomes a key mechanism for creating contrastive supervision that enables discriminative training. 
Conceptually, existing negative sampling strategies can be divided into two families with different levels of modeling sophistication:

\header{Level 0 ($\mathbf{L_0}$): Heuristic-based Negative Sampling.}
This level relies on static, data-driven heuristics independent of the model's training state. Existing heuristic rules typically generate negatives by either treating all unseen items equally~\cite{rendle2012bpr}, leveraging exposure logs to identify unclicked candidates~\cite{ding2019reinforced}, or biasing the sampling distribution according to item popularity~\cite{zhang2019deep, chen2017sampling}.
While highly efficient, these Level 0 strategies primarily produce uninformative negatives. This lack of robust contrastive signals eventually causes gradient vanishing, limiting the model's capacity to learn nuanced preference boundaries under complex data distributions.



\header{Level 1 ($\mathbf{L_1}$): Self-guided Hard Negative Sampling.} To ensure effective parameter updates and compel the model to learn more discriminative user preference representations, hard negative samples are employed—instances difficult to distinguish from positives (e.g., for a positive \texttt{smartphone}, a \texttt{phone case} is more informative than a \texttt{frying pan}). 
To identify hard negative samples, it is standard practice to leverage a \textit{self-guided} sampling mechanism~\cite{huang2020embedding, zhang2013optimizing}, the core logic of which is to identify challenging negative samples in real-time by utilizing the model's \textit{current} parameter state. In practice, this mechanism first constructs a randomly sampled candidate pool, then scores the candidates within this pool based on the active training step's parameters, and finally selects a \textit{single} top-ranked item as the hard negative~\cite{shi2023theories, chen2022learning}. This approach ensures a dynamic alignment between negative sample hardness and the model's evolving predictive capability, thereby facilitating more effective and discriminative parameter updates.

\input{Tables_Algos/DNS+}

\input{Tables_Algos/Advantages}

\header{Remaining Limitations.}
Although L$_1$ methods have achieved certain progress, there still exist the following limitations:

(1) \emph{Negative self-reinforcement loop.} Self-guided hard negative sampling creates a tight coupling between the sampling strategy and the model's current state. This dependency risks triggering a vicious cycle: if the model selects bad negatives, the subsequent gradient updates will degrade the model's discriminative ability. This degradation, in turn, yields even lower-quality samples in future iterations, thereby trapping the model in a local optimum.

(2) \emph{Restricted sampling diversity.} Relying solely on current model parameters for negative sample construction confines the mechanism to select negatives solely within the local item space currently emphasized by the model, failing to effectively explore the broader item spectrum. Consequently, the sampling distribution progressively shrinks, making it difficult to sustain or enhance sample diversity across successive training iterations, thereby severely constraining the model's overall generalization capability.

(3) \emph{Inefficient resource allocation.} The mechanism necessitates scoring the entire pool of potential candidates to identify a single hard negative, leading to significant computational redundancy. The vast majority of the computed scores are discarded without contributing to the gradient update, resulting in a disproportionately high computational overhead relative to the information gain, which severely reduces the overall training efficiency.

\begin{figure}[!t]
  \centering
  \includegraphics[width=1\linewidth]{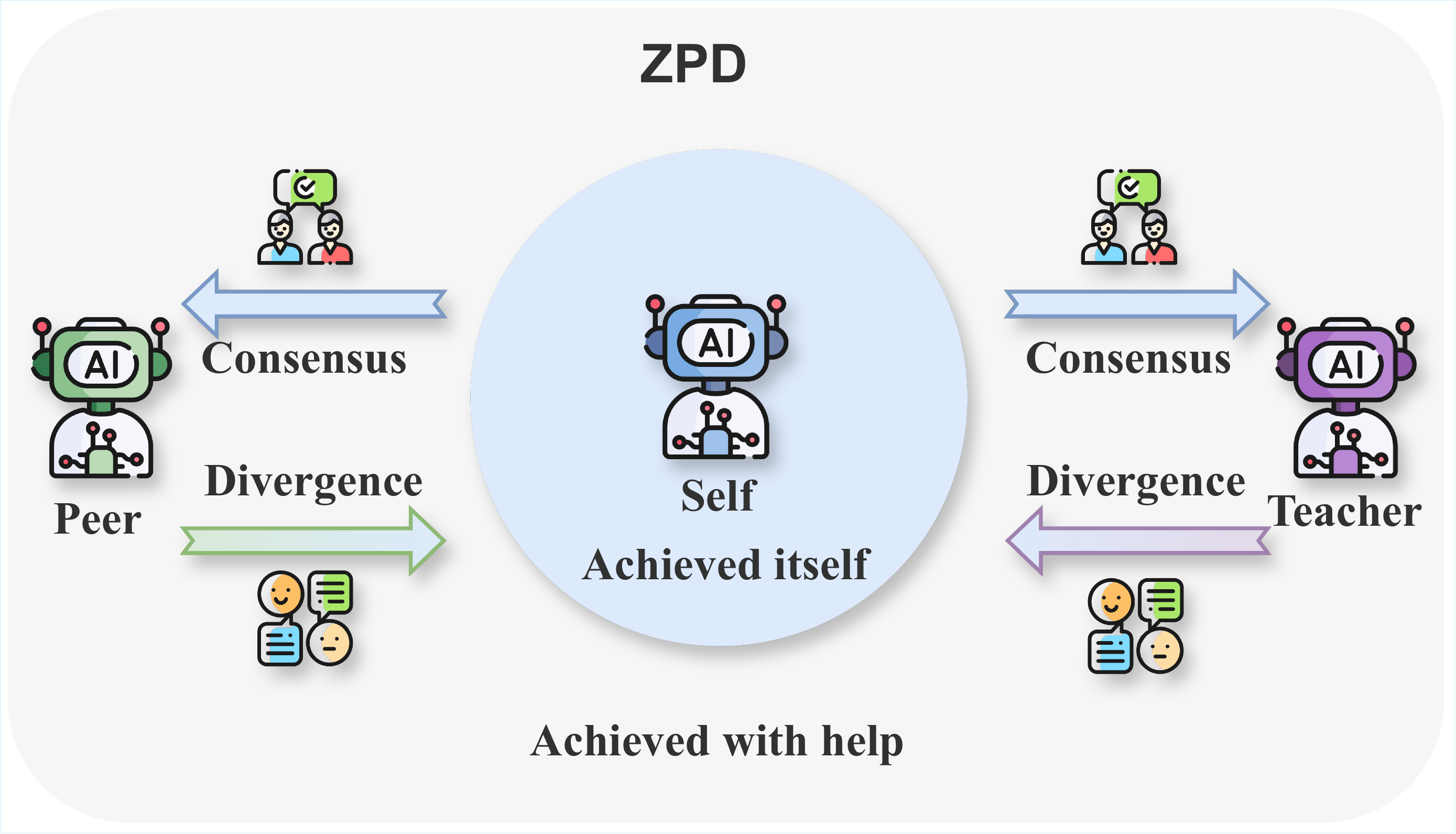}
  \caption{Illustration of the ZPD. It utilizes a "Teacher-Peer-Self" collaborative structure with \textbf{Divergence} and \textbf{Consensus}, thereby extending the learner's capabilities from the inner self-achieved zone to the broader potential zone.}
  \label{fig:zpd}
  \vspace{-1em}
\end{figure}

\header{Our Proposed Framework.} To transcend the inherent limitations of L1-level self-learning, we draw inspiration from Vygotsky’s Zone of Proximal Development (ZPD) theory~\cite{shabani2010vygotsky}. As illustrated in Figure~\ref{fig:zpd}, ZPD posits that optimal learning emerges not in isolation, but within a collaborative triad of "Teacher-Peer-Self". We argue that the engine driving this progress is the dynamic interplay between expanding boundaries (via divergence) and consolidating knowledge (via consensus).
Translating this educational philosophy into the realm of negative sampling, we propose \textbf{MDCNS} (\textbf{\underline{M}}ulti-source \textbf{\underline{D}}ivergence-\textbf{\underline{C}}onsensus for \textbf{\underline{N}}egative \textbf{\underline{S}}ampling). This framework is structurally designed to mirror the ZPD dynamic: it orchestrates multi-source negative samples to maintain a synergy between divergence and consensus, thereby systematically breaking the bottlenecks of isolated sampling methods.

(1) \emph{Multi-source scoring.}
To mitigate the negative self-reinforcement loop, instead of relying solely on the self model, we incorporate a peer model and an ensemble teacher model (derived from the former two) to get candidate items' relevance scores independently. This mechanism disrupts the closed feedback loop, enabling the model to escape suboptimal states and enhancing overall performance.

(2) \emph{Divergence re-ranking.}
To address insufficient sampling diversity, we quantify the score discrepancy between the self model and the peer model across the candidate pool as a divergence score. This term is incorporated into the item relevance scores to re-rank the candidate items. We then randomly select from the top-$M$ re-ranked items to ensure both diversity and appropriate difficulty.

(3) \emph{Consensus distillation.}
To address the problem of inefficient resource allocation, we use the KL divergence~\cite{van2014renyi} to align the score distribution of the self model with that of the teacher model over the candidate item pool. This process facilitates the distillation of consensus knowledge from the teacher to the student (self model), simultaneously maximizing the utilization of training signals.

Table~\ref{tab:related} shows the comprehensive comparison of key characteristics between \model \ and the latest methods.

\begin{figure*}[!t]
  \centering
  \includegraphics[width=1.0\linewidth]{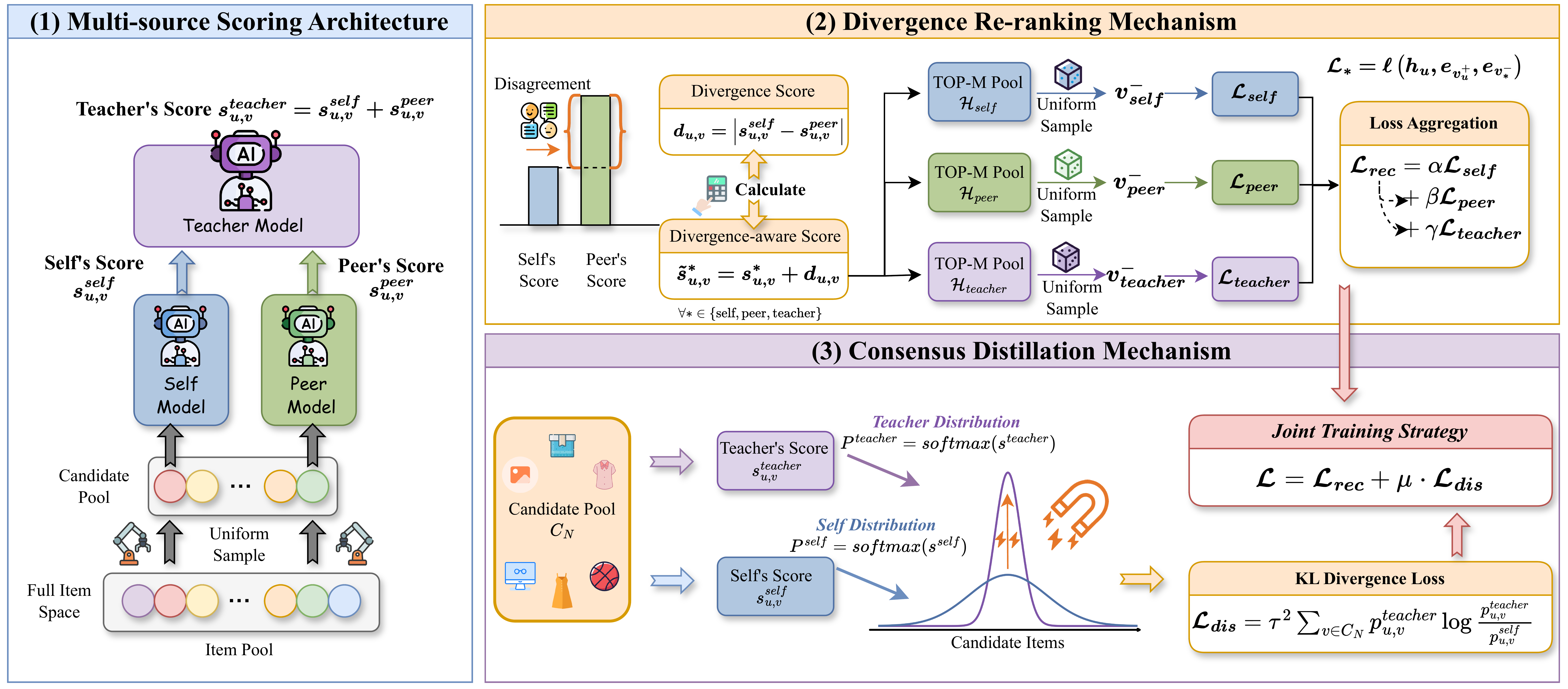}
\caption{The overall framework of \model. 
(1) \textbf{Multi-source Scoring}, which constructs a candidate pool and generates relevance scoring signals from Self, Peer, and Teacher perspectives; 
(2) \textbf{Divergence Re-ranking}, which quantifies view divergence scores between self model and peer model to re-rank candidates and samples diverse hard negatives from the Top-M list; and 
(3) \textbf{Consensus Distillation}, which utilizes the teacher's distribution over the entire candidate pool as soft targets to guide the self model, effectively acquiring consensus knowledge while mitigating the inefficiencies of resource allocation.}
\vspace{-1em}
\label{fig:framework}
\end{figure*}

\header{Contributions.} Our contributions are summarized as follows: 

\begin{itemize}[left=0pt, labelsep=1em] 
\item \emph{Problem Identification.} We identify three critical limitations in existing L$_1$ negative sampling methods, which can be summarized as: the negative self-reinforcement loop, restricted sampling diversity, and inefficient resource allocation.
\item \emph{Theoretical Perspective.} We analyze the bottlenecks of L$_1$ sampling through the lens of Vygotsky’s Zone of Proximal Development (ZPD) theory, proposing a shift from isolated self-guided to a collaborative "Teacher-Peer-Self" paradigm.
\item \emph{Methodological Framework.} We propose \model, a ZPD-inspired framework that technically realizes the "Teacher-Peer-Self" structure via three core components: multi-source scoring, divergence re-ranking, and consensus distillation.
\item \emph{Experimental Validation.} Extensive experiments on various real-world datasets, backbones and loss functions demonstrate that \model\ consistently outperforms state-of-the-art baselines by a significant margin in both effectiveness and generalization.
\end{itemize}






%% file: Tables_Algos/DNS+.tex


%% file: Tables_Algos/Advantages.tex
\begin{table}[t]
    \centering
    \small
\caption{
Characteristics comparison between \model \ and existing methods. The abbreviations denote: \textbf{SGS} (Self-Guided Sampling), \textbf{MSS} (Multi-Source Scoring), and \textbf{ERA} (Efficient Resource Allocation). In the PRF column, a higher number of \oca \ marks indicates superior performance.
}
    \setlength{\tabcolsep}{6pt} 
    \renewcommand{\arraystretch}{1.0} 
    \begin{tabularx}{0.45\textwidth}{l *{5}{>{\centering\arraybackslash}X}} 
      \toprule
      \textbf{Level} &
      \textbf{Method} 
      & \textbf{SGS} & \textbf{MSS} & \textbf{ERA}& \textbf{PRF} \\
      \midrule
      \multirow{3}{*}{$\mathbf{L_0}$}&
      RNS~\cite{rendle2012bpr}
        & \no &\no &\yes& \oca \\
      &PNS~\cite{chen2017sampling}
        & \no &\no &\yes& \oca \\
      &ENS~\cite{ding2019reinforced}
        & \no &\no &\yes& \oca \\
        \midrule
      \multirow{6}{*}{$\mathbf{L_1}$}&
            DNS~\cite{zhang2013optimizing}
        & \yes &\no &\no& \oca\oca \\
      &MixGCF~\cite{huang2021mixgcf}
        & \yes&\no &\no& \oca\oca\\
      &AdaSIR~\cite{chen2022learning}
        & \yes&\no &\no& \oca\oca\\
      &GNNO~\cite{fan2023neighborhood} 
        & \yes & \no & \no & \oca\oca\\
        &DNS+~\cite{shi2023theories}
        & \yes & \no & \no & \oca\oca\\
&SRNS~\cite{ding2020simplify}& \yes & \no &\no& \oca\oca\\
\midrule
      \multirow{1}{*}{$\mathbf{L_2}$}&
      \textbf{\model} 
        & \textcolor{green!60!black}{\ding{51}} & \textcolor{green!60!black}{\ding{51}} & \textcolor{green!60!black}{\ding{51}}& \oca\oca\oca\\
      \bottomrule
    \end{tabularx}

    \label{tab:related}
  \vspace{-1em}
    
\end{table}

%% file: sections/P2.tex
\section{Problem Formulation}
\label{sec:Problem}

Let $\mathcal{U}$ and $\mathcal{V}$ denote the sets of users and items, respectively. We represent the interaction history of a user $u \in \mathcal{U}$ as a chronologically ordered sequence $S_{u}=[v_{1}, v_{2}, \ldots, v_{|S_u|}]$, where $v_{t} \in \mathcal{V}$ indicates the item interacted with at time step $t$.

The primary objective of sequential recommendation is to forecast the user's next interaction at step $|S_u|+1$. Formally, we aim to identify the candidate item $v^*$ that maximizes the conditional probability given the user's history:
\begin{equation}
    \underset{v^{*} \in \mathcal{V}}{\arg \max }\;\; P\left(v_{|S_u|+1} =v^{*} \mid S_{u} \right).
\end{equation}
\vspace{-1pt}
To optimize for this objective, state-of-the-art methods typically employ either a pairwise ranking loss, such as bayesian personalized ranking $\mathcal{L}_{bpr}$, or a binary cross-entropy loss $\mathcal{L}_{bce}$. Specifically, $\mathcal{L}_{bpr}$ is designed to learn a scoring function that ranks the ground-truth next item (positive sample, $v^+$) higher than unobserved items (negative samples, $v^-$). Conversely, $\mathcal{L}_{bce}$ formulates the recommendation task as a binary classification problem, distinguishing positive instances $v^+$ from sampled negative instances $v^-$.

%% file: sections/method.tex
\vspace{-0.6em}
\section{Methodology}
In this section, motivated by the Zone of Proximal Development (ZPD) theory, we propose \model, a novel "Teacher-Peer-Self" negative sampling framework for sequential recommendation. Specifically, \model\  incorporates three synergistic components: multi-source scoring, divergence re-ranking, and consensus distillation.

As illustrated in Figure~\ref{fig:framework}, \model \ first employs a multi-source scoring phase (Section~\ref{sec:Multi-source sampling}) to derive candidate item relevance scores from the distinct perspectives of the teacher, peer, and self.
Then, in the divergence re-ranking phase (Section~\ref{sec:Divergence Re-ranking}), the divergence score is computed between the predictions of the peer and self models. Subsequently, the candidates are re-ranked by jointly considering both the item relevance scores and the divergence scores.
Furthermore, a consensus distillation module (Section~\ref{sec:Consensus Distillation}) is proposed to fully exploit the computational cost and distill the consensus knowledge from the teacher model into the self model.
Finally, the time complexity of the proposed \model \ is analyzed in Section~\ref{sec:timecost}.

\vspace{-0.6em}
\subsection{Multi-source Scoring}\label{sec:Multi-source sampling}
Formally, let $S_{u}=[v_{1}, v_{2}, \ldots, v_{|S_u|}]$ denote the interaction sequence for user ${u}$, where ${v_t}$ represents the item interacted at step $t$.
In our framework, we denote the self model to be trained as ${M_{self}}$ (e.g., SASRec) and the peer model as ${M_{peer}}$ (e.g., Mamba4Rec)\footnote{$M_{self}$ and $M_{peer}$ function as mutual peers within a dual framework. $M_{peer}$ follows an identical training procedure to $M_{self}$, we omit redundant descriptions for brevity.}, and the teacher model as ${M_{teacher}}$.

\subsubsection{Candidate Pool Construction.}
Given the computational intractability of performing full-ranking over the vast item space $\mathcal{V}$, we adopt a candidate generation strategy. Specifically, we construct a candidate set $C_N$ by uniformly sampling $N$ items from the unobserved item space (i.e., excluding the user's history $\mathrm{S}_u$ and the ground-truth target $v_u^+$). Formally, $C_N$ is obtained via:

\begin{equation}\label{eq:2}
    C_N \sim \text{Uniform} \left( \mathcal{V} \setminus \left( \{ v_u^+ \} \cup {S}_u \right), N \right).
\end{equation}

\subsubsection{Dual-View Scoring.}
Both models encode the input sequence ${S_u}$ into latent representations. Let ${h_u^{self}}$ and ${h_u^{peer}}$ denote the sequence representations generated by ${M_{self}}$ and ${M_{peer}}$, respectively. For every candidate item ${v} \in {C_N}$, let ${e_v}$ be its item embedding. We compute the relevance scores by performing the dot product between the sequence representation and the item embeddings:
\begin{equation}\label{eq:3}
    {s_{u,v}^{self}} = {h_u^{self}} \cdot {e_v}, \quad {s_{u,v}^{peer}} = {h_u^{peer}} \cdot {e_v},
\end{equation}
where ${s_{u,v}^{self}}$ and ${s_{u,v}^{peer}}$ represent the prediction scores for item ${v}$ under the self and peer views, respectively. Consequently, we obtain the score sets for the candidate items: ${S_{C_N}^{self}} = \{ {s_{u,v}^{self}} \mid {v} \in {C_N} \}$ and ${S_{C_N}^{peer}} = \{ {s_{u,v}^{peer}} \mid {v} \in {C_N} \}$.

\subsubsection{Teacher Signal Aggregation.}
To mitigate the bias inherent in a single model and leverage the consensus knowledge, we define the teacher model ${M_{teacher}}$ as an additive aggregation of the two student models. The teacher's score for each candidate item ${v} \in {C_N}$ is calculated as:
\begin{equation}\label{eq:4}
    {s_{u,v}^{teacher}} = {s_{u,v}^{self}} + {s_{u,v}^{peer}}.
\end{equation}
We utilize ${S_{C_N}^{teacher}}$ for two distinct purposes: guiding negative sampling from the teacher's perspective and serving as soft targets for distilling knowledge into the self model.

\subsection{Divergence Re-ranking}\label{sec:Divergence Re-ranking}
To alleviate the issue of restricted sampling diversity, we propose a divergence-based re-ranking strategy. This approach exploits the disagreement between the self and peer views to identify samples that are not only hard but also diverse.

\subsubsection{Divergence Quantification and Re-scoring.}
We hypothesize that items with high scoring discrepancies between the self and peer models represent uncertain samples. For each candidate item ${v} \in {C_N}$, we quantify this disagreement as the absolute difference between their prediction scores, termed the Divergence Score ${d_{u,v}}$:
\begin{equation}\label{eq:5}
    d_{u,v} = \left| s_{u,v}^{self} - s_{u,v}^{peer} \right|.
\end{equation}
To prioritize these informative samples, we augment the original prediction scores with the divergence term. Specifically, we update the scores for the self, peer, and teacher views as follows:
\begin{equation}\label{eq:6}
    \tilde{s}_{u,v}^{\ast} = s_{u,v}^{\ast} + d_{u,v}, \quad \forall \ast \in \{ \text{self}, \text{peer}, \text{teacher} \}
\end{equation}
where $\tilde{s}_{u,v}^{\ast}$ denotes the divergence-aware score. This formulation ensures that items with both high relevance probability and high view disagreement are ranked higher.

\subsubsection{Hard Negative Sampling.}
To construct robust negative sets, we first identify the most informative candidates for each view. Specifically, for each perspective ${\ast} \in \{ \text{self}, \text{peer}, \text{teacher} \}$, we select the top-${M}$ items from the candidate pool ${C_N}$ based on their divergence-aware scores $\{ \tilde{s}_{u,k}^{\ast} \}$:
\begin{equation}\label{eq:7}
    \mathcal{H}_{\ast} = \operatorname{Top-M}\left( \{ \tilde{s}_{u,k}^{\ast} \mid k \in C_N \} \right).
\end{equation}
Subsequently, to mitigate the false negative problem inherent in strict hard mining and to introduce stochasticity for robust training, we uniformly sample one negative instance from each top-ranked pool:
\begin{equation}\label{eq:8}
    v_{\ast}^- \sim \mathrm{Uniform}\left( \mathcal{H}_{\ast} \right).
\end{equation}
In this manner, the negative samples are selected to ensure that they are not only challenging but also diverse.

\subsubsection{Recommendation Optimization Objective.}
To train the Self-Model $M_{self}$, we employ a standard ranking loss $\ell(\cdot)$ (e.g., BPR or BCE) and formulate the view-specific optimization objective as:
\begin{equation}\label{eq:9}
\mathcal{L}{\ast} = \ell \left( h_u^{self}, e_{v_{u}^+}, e_{v_{\ast}^-} \right),
\end{equation}
where ${h_u^{self}}$ is the user representation from the self-model, ${e_{v_u^+}}$ is the embedding of the ground-truth positive item, and ${e_{v_{\ast}^-}}$ is the embedding of the negative item sampled from the perspective ${\ast} \in \{ \text{self}, \text{peer}, \text{teacher} \}$. 
Finally, the total recommendation objective is computed as a weighted aggregation of the losses derived from the three distinct negative sampling distributions:
\begin{equation}\label{eq:10}
    \mathcal{L}_{rec} = \alpha \mathcal{L}_{self} + \beta \mathcal{L}_{peer} + \gamma \mathcal{L}_{teacher},
\end{equation}
where $\alpha$, $\beta$, and $\gamma$ are hyperparameters that balance the contribution of hard negatives mined from the views of the self, the peer and the teacher, respectively.

\input{Tables_Algos/R2NS_algo}

\subsection{Consensus Distillation}\label{sec:Consensus Distillation}
The previous stage incurs the computational cost of scoring the candidate set $C_N$, yet standard sampling exploits only one instance. To maximize the utility of these computations, we propose a consensus distillation module. By distilling the distribution of $M_{teacher}$ over $C_N$ into $M_{self}$, we preserve the global ranking signals that are otherwise discarded by hard sampling.
\subsubsection{Probabilistic Soft Labeling.}
To align the granular ranking views of the teacher and self-model, we transform their raw scores into probability distributions over the candidate set $C_N$. Specifically, we employ a temperature-scaled Softmax function to normalize the scores. For a user $u$ and item $v \in C_N$, the probability under view $\ast \in \{ \text{self}, \text{teacher} \}$ is computed as:
\begin{equation}\label{eq:11}
p_{u,v}^{\ast} = \frac{\exp(s_{u,v}^{\ast} / \tau)}{\sum_{j \in C_N} \exp(s_{u,j}^{\ast} / \tau)},
\end{equation}
where $\tau$ is a temperature hyperparameter regulating the sharpness of the distribution.

\subsubsection{Distillation Objective.}
We adopt the KL Divergence~\cite{van2014renyi} to minimize the discrepancy between the teacher's consensus distribution and the self-model's predicted distribution. Following the standard knowledge distillation paradigm \cite{hinton2015distilling}, we use the temperature $\tau$ as the scaling factor to control the gradient magnitudes. The distillation loss is formulated as:
\begin{equation}\label{eq:12}
    \mathcal{L}_{dis} = \tau^2 \sum_{v \in C_N} p_{u,v}^{teacher} \log \frac{p_{u,v}^{teacher}}{p_{u,v}^{self}}.
\end{equation}
 By minimizing $\mathcal{L}_{dis}$, ${M_{self}}$ is encouraged to approximate the ensemble prediction ${M_{teacher}}$, thereby implicitly incorporating the teacher knowledge and improving generalization robustness.

\subsubsection{Joint Training Strategy.}
To simultaneously optimize for accurate ranking of the ground-truth items and alignment with the ensemble consensus, we train the framework in a multi-task learning manner. The final objective function combines the recommendation loss derived from multi-source hard negatives (Equation~\ref{eq:9}) and the consensus distillation loss:
\begin{equation}\label{eq:13}
    \mathcal{L} = \mathcal{L}_{rec} + \mu \cdot \mathcal{L}_{dis},
\end{equation}
where $\mu$ is a weighting coefficient that controls the strength of the distillation signal. By jointly optimizing this objective, our framework ensures that the self-model ${M_{self}}$ not only discriminates the positive item from hard negatives effectively but also maintains a holistic understanding of the global preference distribution.

\subsection{Time Complexity Analysis}\label{sec:timecost}


We analyze the time complexity of the proposed framework across its three distinct stages. 
In the \textit{multi-source scoring} stage, constructing the candidate pool requires $\mathcal{O}(N)$, while dual-view scoring and teacher signal aggregation incur $\mathcal{O}(Nd)$ and $\mathcal{O}(N)$, respectively. Consequently, the complexity of this stage is bounded by $\mathcal{O}(Nd)$. 
In the \textit{divergence re-ranking} stage, computing divergence-aware scores operates in $\mathcal{O}(N)$. The hard negative sampling procedure introduces a term of $\mathcal{O}(N \log M)$, where $M$ denotes the size of the random sampling pool. Since the optimization objective cost $\mathcal{O}(d)$ is negligible compared to the sampling cost (i.e., $d \ll N \log M$), this stage scales as $\mathcal{O}(N \log M)$. 
Finally, the \textit{consensus distillation} stage involves probabilistic soft labeling and KL distillation, both of which operate in $\mathcal{O}(N)$. 
Comparing the costs across stages, and given that $\log M \ll d$, the $\mathcal{O}(N \log M)$ term from the re-ranking stage is dominated by the $\mathcal{O}(Nd)$ term from the scoring stage. Therefore, the time complexity of the entire framework is $\mathcal{O}(Nd)$.

This is on par with representative state-of-the-art baselines such as DNS+ and MixGCF, whose complexities are also $\mathcal{O}(Nd)$. 
Therefore, our method does not incur additional asymptotic computational overhead, while fully utilizing the available computational cost to deliver improved performance.


%% file: Tables_Algos/R2NS_algo.tex

\begin{algorithm}[!t]
    \SetAlgoLined
    \DontPrintSemicolon  
    \KwIn{Interaction sequences, candidate size $N$}
    \KwOut{Learned parameter $\Theta$ of $M_{\text{self}}$}
    
    \For{$epoch \gets 1$ \KwTo $T$}{
        \ForEach{sequence $S_u$}{
            \textcolor{gray}{\textbf{// Multi-source Scoring}}\;
            Construct candidate set $C_N$ according to Eq.~(\ref{eq:2})\;
            Obtain $s^{\text{self}}, s^{\text{peer}}$ and $s^{\text{teacher}}$ for $C_N$ (Eqs.~(\ref{eq:3})--(\ref{eq:4}))\;
            
            \vspace{0.2em} 
            \textcolor{gray}{\textbf{// Divergence Re-ranking}}\;
            Get divergence-aware scores $\tilde{s}$ using Eq.~(\ref{eq:5}) to Eq.~(\ref{eq:6})\;
            \ForEach{view $\ast \in \{ \text{self}, \text{peer}, \text{teacher} \}$}{
                Select $v_{\ast}^{-}$ from Top-M candidates (Eqs.~(\ref{eq:7})--(\ref{eq:8}))\;
            }
            Calculate ranking loss $\mathcal{L}_{rec}$ via Eqs.~(\ref{eq:9})--(\ref{eq:10})\;
            
            \vspace{0.2em}
            \textcolor{gray}{\textbf{// Consensus Distillation}}\;
            Calculate consensus loss $\mathcal{L}_{dis}$ via Eqs.~(\ref{eq:11})--(\ref{eq:12})\;
            Update $\Theta$ by minimizing $\mathcal{L}_{rec} + \mu \mathcal{L}_{dis}$\;
        }
    }
    \Return{$M_{\text{self}}$ with $\Theta$}
    \caption{The training process of \model}
    \label{alg:method}
\end{algorithm}

\vspace{-1em}

%% file: sections/experiments.tex
\input{Tables_Algos/data_statitcs}

\input{Tables_Algos/overall_performance}

\section{Experiments}
\label{sec:experiments}
In this work, we aim to answer the following research questions:
\begin{enumerate}[leftmargin=*, label=\textbf{(RQ\arabic*)}]
     \item How does \model\ perform compared to existing negative sampling methods in sequential recommendation?
    \item How well does \model\ generalize across recommendation models, loss functions and weak to strong setting?
    \item How do components of \model\  affect the performance?
    \item How does the loss of \model\ evolve during training?
\end{enumerate}
\subsection{Experiments Setup}\label{sec:exp_setup}
\label{subsection:datasset}
\subsubsection{Datasets}
Experiments are conducted on six real-world datasets: \textbf{Sports, Beauty, Toys} and \textbf{Health} are derived from Amazon product reviews~\cite{mcauley2015image}, corresponding to the "Sports and Outdoors", "Beauty", "Toys and Games" and "Health and Personal Care" categories, respectively. \textbf{KuaiRand}~\cite{gao2022kuairand} is retrieved from the recommendation logs of the video-sharing mobile application Kuaishou. It notably contains millions of user interactions with randomly exposed items.
\textbf{LastFM}\footnote{\url{https://grouplens.org/datasets/hetrec-2011/}} is a music artist recommendation dataset derived from user tagging behaviors. 
In this work, the user-assigned tags associated with artists are utilized as item attributes.
For consistency with prior work~\cite{zhou2020s3,kang2018self}, we process the Amazon datasets by retaining only the 5-core filtered version where each item/user appears in at least five interactions. The KuaiRand and LastFM undergo identical preprocessing. Table~\ref{tab:datasets} presents a comprehensive summary of the key statistics for our six processed datasets.

\subsubsection{Evaluation Proposal}
\label{subsec:Evaluation proposal}
In alignment with established protocols~\cite{zhou2020s3, kang2018self}, we assess model performance utilizing Recall@$K$ and NDCG@$K$ across $K \in \{5, 10, 20\}$. 
Regarding dataset partitioning, we implement a ratio splitting scheme, allocating 70\% of sequences for training, 20\% for validation, and the remaining 10\% for testing.

\input{Tables_Algos/model_roub}

\input{Tables_Algos/loss_roub}

\subsubsection{Backbone Models}\label{sec:backbone_models} 
The method proposed herein is inherently model-agnostic, enabling seamless integration into the training processes of a wide range of sequential recommendation architectures. To evaluate its generalization ability and effectiveness, we conduct a series of extensive experiments using several representative backbone models from diverse architectural paradigms.
\begin{itemize}[leftmargin=*] 
\item \textbf{MLP-based}: \textbf{FMLP4Rec}~\cite{zhou2022filter} represents an all-MLP architecture that encodes sequential dependencies via learnable filters. 
\item \textbf{RNN-based}: \textbf{GRU4Rec}~\cite{hidasi2015session} uses Gated Recurrent Units to model the temporal evolution of user interests effectively. 
\item \textbf{Transformer-based}: \textbf{SASRec}~\cite{kang2018self} leverages multi-head self-attention to capture long-range dependencies within sequences. 
\item \textbf{Linear Transformer-based}: \textbf{LinRec}~\cite{liu2023linrec} adopts a linearized attention mechanism for efficient sequence modeling.
\item \textbf{State Space Model-based}: \textbf{Mamba4Rec}~\cite{liu2024mamba4rec} applies the Selective State Space Model to achieve efficient sequence processing.
\end{itemize}

\vspace{-3pt}
\subsubsection{Baselines} 
We compare \model\  with various state-of-the-art hard negative-focused methods and false negative mitigated methods on sequential recommendation models:
\begin{itemize}[leftmargin=*]
\item \textbf{RNS}~\cite{rendle2012bpr}  employs a uniform distribution to select negatives.
\item \textbf{Hard negative-focused methods:}
  \begin{enumerate}[leftmargin=*]
\item \textbf{DNS}~\cite{zhang2013optimizing} is a dynamic hard negative sampling technique, where negatives are chosen based on the highest item scores.
\item \textbf{MixGCF}~\cite{huang2021mixgcf} generates hard negatives by mixing positive signals from the interaction graph into negative samples through both positive and hop mixing strategies.
\item \textbf{AdaSIR}~\cite{chen2022learning} employs a two-stage process that maintains a sample pool, with importance resampling to select negatives.
  \end{enumerate}
 \item \textbf{False negative-mitigated methods:}
\begin{enumerate}[leftmargin=*]
\item \textbf{GNNO}~\cite{fan2023neighborhood} identifies negatives based on neighborhood overlap and mitigates false negatives by evaluating similarity.
\item \textbf{DNS+}~\cite{shi2023theories} enhances DNS by randomly sampling the top-$M$ candidates to reduce the occurrence of false negatives.
\item \textbf{MixGCF+}~\cite{huang2021mixgcf} enhances MixGCF by randomly sampling the top-$M$ candidates to reduce the occurrence of false negatives.
\item \textbf{SRNS}~\cite{ding2020simplify} employs a score-based memory and variance-based sampling technique to ensure high-quality negative samples.
\end{enumerate}
\end{itemize}


\subsection{Implementation Details}\label{sec:implementation}

We adopt the official RecBole implementations~\cite{zhao2022recbole, zhao2021recbole} for all backbone models and follow the hyperparameter configurations recommended in the original papers whenever applicable. For negative sampling baselines, we use the released implementations and hyperparameter search spaces reported in their respective works to ensure a fair comparison. All models are optimized with Adam~\cite{kingma2014adam} (learning rate $lr = 0.001$, $\beta_1 = 0.9$, $\beta_2 = 0.999$), using an embedding size of 64 and a batch size of 1024.
We set $N = 100$, and tune $M$ over the range $\{1, 2, 3, 4, 5\}$. The balance coefficients $\alpha$, $\beta$, and $\gamma$ are selected via grid search in the interval $[0.5, 3.0]$ with a step size of $0.5$, while the temperature $\tau$ is chosen from the set $\{0.7, 1.0, 10.0\}$. The coefficient $\mu$ is fixed to $1$. Early stopping on the validation set with a patience of $30$ is used to ensure training convergence. During evaluation, we adopt a full item ranking protocol and use only the self model $M_{\text{self}}$ to generate recommendations. To mitigate random variance, we report the final results on the test set as the average over three independent runs.

\input{Tables_Algos/ablation}

\input{Tables_Algos/WTS}

\subsection{Performance of \model\ (RQ1)}\label{sec:rq1}
The comprehensive performance evaluation contrasting \model\ against diverse baseline negative sampling frameworks across six datasets is detailed in Table~\ref{tab:overall}.
For the experimental configuration, the foundational SASRec is employed as the self model, accompanied by Mamba4Rec acting as the peer model. Both architectures are optimized using the standard binary cross-entropy (BCE) loss~\cite{shannon1948mathematical}.
Empirical findings indicate that the introduced \model\ reliably achieves superior performance across all evaluated metrics and data domains.
Most impressively, \model\ registers significant enhancements over the most competitive baselines, peaking at a 27.29\% improvement in Recall@20 and a 36.44\% surge in NDCG@20 on the Beauty dataset. 
It is also particularly noteworthy that the framework consistently yields state-of-the-art results on non-Amazon datasets (namely, LastFM and KuaiRand), thereby demonstrating exceptional domain generalization and data robustness.

\textbf{(Answer to RQ1)} \model\ systematically outperforms existing negative sampling methods across diverse datasets, achieving state-of-the-art performance with substantial metric enhancements.

\subsection{Generalization Evaluation (RQ2)}\label{sec:rq2}
In this section, we demonstrate the generalization abilities of \model\  through experiments on different recommendation backbone models, various loss functions and weak to strong setting.
\subsubsection{Generalization across Different Backbone Models.}\label{sec:exp_model}
The proposed \model\ is designed as a model-agnostic negative sampling framework, offering high flexibility across various sequential recommendation backbones. 
To validate this plug-and-play capability, the framework is deployed across a diverse suite of five backbone models, following the details in Section~\ref{sec:backbone_models}. In these configurations, Mamba4Rec generally serves as the peer model, with the exception of the Mamba4Rec backbone evaluation, where SASRec is utilized in this role. 
As evidenced by the results in Tables~\ref{tab:model_rob} and \ref{tab:loss_rob}, \model\ achieves significant and statistically robust performance gains across all tested backbones, consistently outperforming established baselines. 
Notably, \model\ demonstrates distinct advantages across a spectrum of architectures, ranging from filter MLP models (e.g., FMLP4Rec) to advanced mamba models (e.g., Mamba4Rec). The substantial improvements observed even on relatively simpler RNN based GRU4Rec further underscore its exceptional robustness. 
These compelling results substantiate that \model\ can be universally and reliably applied to enhance the performance of a wide range of sequential recommendation systems.

\subsubsection{Generalization Across Varied Loss Functions.}\label{sec:exp_loss}
In addition to its architectural flexibility, \model\ demonstrates strong performance across a range of optimization objectives. 
To explore this versatility, we evaluate \model\ using two widely-used loss functions: Binary Cross-Entropy (BCE) loss~\cite{shannon1948mathematical}, which treats recommendation as a classification problem, and Bayesian Personalized Ranking (BPR) loss~\cite{rendle2012bpr}, which focuses on optimizing pairwise rankings. 
As detailed in Tables~\ref{tab:model_rob} and~\ref{tab:loss_rob}, \model\ consistently outperforms baseline sampling methods across various datasets and evaluation metrics under both loss functions. 
This highlights that the performance gains with \model\ are largely independent of the chosen optimization objective. 
These results affirm the robustness and flexibility of \model\ across different learning paradigms, underlining its ability to generalize effectively and adapt to a wide range of recommendation tasks. 
This positions \model\ as a highly versatile tool, well-suited for sequential recommendation across diverse settings.

\begin{figure*}[!t]
  \centering
  \includegraphics[width=1\linewidth]{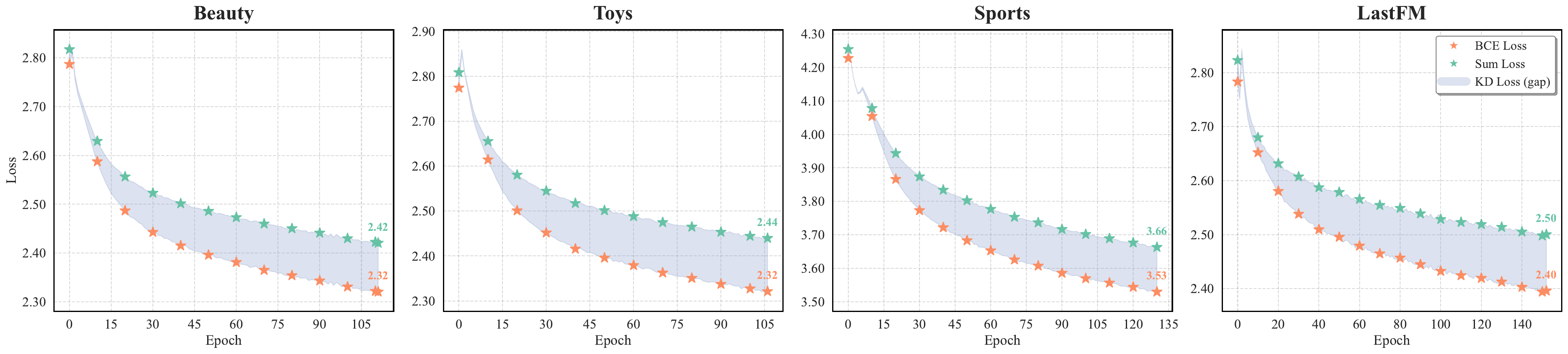}
  \caption{Loss dynamics of \model \ on four datasets.}
  \vspace{-1em}
  
  \label{fig:loss}
\end{figure*}

\subsubsection{Generalization across Weak to Strong Setting.}
To further assess the robustness of our framework in a "weak-to-strong" generalization paradigm, we examine the influence of a weaker peer model on overall performance. 
Specifically, we use SASRec as the backbone and Mamba4Rec as the peer model, denoted as \model, and compare it with $\text{\model}_{\text{wts}}$, where SASRec is the backbone and a weak model GRU4Rec serves as the peer model.
As shown in Table~\ref{tab:wts}, the performance dynamics follow a dualistic pattern. 
On the Beauty and Toys datasets, the standard model \model\  outperforms $\text{\model}_{\text{wts}}$. 
However, on the Sports and LastFM datasets, $\text{\model}_{\text{wts}}$ achieves competitive results, even marginally outperforming \model. 
Importantly, both variants consistently show substantial improvements over the baselines, regardless of these dataset-specific variations. 
This highlights the structural robustness of our framework, demonstrating that its effectiveness is not strictly dependent on utilizing a state-of-the-art peer model. 
We attribute this resilience to our negative sampling mechanism, which disrupts the negative self-reinforcement loop and introduces essential diversity and difficulty by divergence re-ranking, even when signals are derived from a weaker model.

\textbf{(Answer to RQ2)} \model\ exhibits exceptional generalization, consistently delivering robust performance gains across diverse backbones, varied loss functions, and weak-to-strong settings.

\vspace{-1em}
\subsection{Ablation Study (RQ3)}\label{sec:rq3}
In this section, we perform a series of ablation studies to examine the contribution of each component in \model, namely multi-source scoring (MS), divergence re-ranking (DR), and consensus distillation (CD). All experiments are conducted on the SASRec backbone model, using three Amazon datasets. The results are in Table~\ref{tab:ablation}. The following variants are evaluated in our study:

\begin{itemize}[leftmargin=*]
    \item \textbf{w/o CD}: This variant excludes consensus distillation and directly utilizes the recommendation loss, $\mathcal{L}_{rec}$, for model training.
    \item \textbf{w/o CD-DR}: In this variant, divergence re-ranking is removed, meaning the score differences between \( M_{\text{self}} \) and \( M_{\text{peer}} \) in the candidate set are not used. Instead, hard negative samples are selected using the original scores.
    \item \textbf{w/o CD-DR-MS}: This variant omits multi-source negative scoring, relying exclusively on the scores from $M_{\text{self}}$ to mine hard negative samples, without incorporating negative samples from other sources, which degenerates to the base DNS+ method.
\end{itemize}

As shown in Table~\ref{tab:ablation}, \model\ consistently outperforms all ablation variants. Removing the consensus distillation phase ("w/o CD") results in a clear performance drop, highlighting the importance of extracting information from the teacher's consensus.
Further exclusion of divergence re-ranking ("w/o CD-DR") leads to noticeable degradation, confirming its effectiveness in selecting more diverse negative samples.
The additional decline observed in the absence of multi-source sampling ("w/o CD-DR-MS") further underscores the crucial role of multi-source negative sampling in breaking the negative self-reinforcement loop.

\textbf{(Answer to RQ3)} The progressive performance degradation across all ablation variants confirms that multi-source scoring, divergence re-ranking, and consensus distillation are all mutually indispensable for the optimal performance of \model.

\subsection{Training Dynamics Analysis}
To empirically validate the stability and convergence of our optimization process, we analyze the training loss trajectories visualized in Figure~\ref{fig:loss}.
We observe that both the BCE recommendation loss and the cumulative sum loss exhibit a steady, monotonic decrement, eventually stabilizing at a minimal bound. This convergence behavior attests to the robustness of our proposed framework.
A pivotal observation lies in the evolution of the knowledge distillation loss, represented by the gap between the curves. Unlike conventional distillation where the gap often diminishes rapidly, our KD loss maintains a significant magnitude or expands throughout the training phase. 
We attribute this phenomenon to the proposed divergence re-ranking mechanism. 
By actively preventing the teacher and self-models from collapsing into homogenized representations, this mechanism ensures that the teacher continuously provides an informative gradient for the student.
This sustained knowledge transfer capability further corroborates the effectiveness of our approach from the perspective of optimization dynamics.

\textbf{(Answer to RQ4)} \model\ exhibits robust training dynamics where primary losses converge steadily, yet distillation loss expands to provide sustained informative gradients for optimization.


%% file: Tables_Algos/data_statitcs.tex
\begin{table}[!t]
	\centering
  \caption{Dataset statistics.}
  \vspace{-1em}
        \renewcommand\arraystretch{0.95}
	\scalebox{0.85}{
		\begin{tabular}{l!{\vrule}rrrrrr}
			\toprule \textbf{{Dataset}} & \textbf{Health} & \textbf{Sports} &\textbf{Beauty} & \textbf{Toys} & \textbf{LastFM} & \textbf{KuaiRand} \\
			\midrule 
			{\# Users} &66,519 & 35,598 & 22,363 &  19,412&1,090& 18,642\\
			 {\# Items} &28,237  & 18,357 & 12,101  & 11,924&3,646&7,566 \\
			 {\# Inter} & 485,163 & 296,337 & 198,502 & 167,597& 52,551&231,471\\
              {\# AvgLen} &7.3  & 8.3 & 8.8  & 8.6& 48.2& 12.4\\
			 {Sparsity} & 99.97\% & 99.95\% & 99.93\% & 99.93\%& 98.68\%& 99.83\%\\
			\bottomrule
	\end{tabular}}
	\label{tab:datasets}
\end{table}

%% file: Tables_Algos/overall_performance.tex
\begin{table*}[!t]
  \centering
  \caption{Performance comparison across six datasets. The best and the second-best scores are marked in bold and \underline{underlined} fonts. RC and NG denote Recall and NDCG. * denotes p-value < 0.05 for paired t-tests.}
  \vspace{-1em}
  \begin{tabular}{c!{\vrule}ccc!{\vrule}ccc!{\vrule}ccc!{\vrule}ccc}
  \toprule
  \multirow{2}{*}{Method} & \multicolumn{6}{c!{\vrule}}{Beauty} & \multicolumn{6}{c}{Toys} \\
   & RC@5 & RC@10 & RC@20 & NG@5 & NG@10 & NG@20 & RC@5 & RC@10 & RC@20 & NG@5 & NG@10 & NG@20 \\
  \midrule
  RNS    & 0.0434 & 0.0612 & 0.0966 & 0.0289 & 0.0347 & 0.0435
  & 0.0592 & 0.0865 & 0.1231 & 0.0406 & 0.0496 & 0.0587 \\
 DNS &0.0398 & 0.0545 & 0.0715 & 0.0288 &  0.0335 & 0.0378 & 0.0216 & 0.0288 & 0.0366  & 0.0170 & 0.0193 & 0.0213 \\
 MixGCF & 0.0349& 0.0550 & 0.0805 & 0.0232 & 0.0297 & 0.0361 &0.0144 & 0.0175 & 0.0237 & 0.0106 & 0.0117 & 0.0132 \\
 
 AdaSIR &\underline{0.0590}&0.0840&0.1127&\underline{0.0417}&\underline{0.0497}&\underline{0.0568}&0.0685&0.0947&0.1241&0.0487&0.0573&0.0647\\
\hdashline
  GNNO       & 0.0577 & \underline{0.0854} & 0.1144 & 0.0388 & 0.0476 & 0.0550 
  & 0.0705 & \underline{0.0989} & 0.1292 & 0.0475 & 0.0566 & 0.0641\\

DNS+       & 0.0550 & 0.0769 & 0.1140 & 0.0355 & 0.0425 & 0.0519 
  & 0.0721 & 0.0963 & \underline{0.1344} & 0.0503 & 0.0581 & \underline{0.0676} \\
 MixGCF+     & 0.0532 & 0.0720 & 0.1068 & 0.0359 & 0.0418 & 0.0506
  & 0.0685 & 0.0968 & 0.1282 & 0.0463 & 0.0554 & 0.0635\\ 
SRNS &0.0487&0.0742&\underline{0.1180}&0.0334&0.0416&0.0527&\underline{0.0772}&0.0978&0.1308&\underline{0.0521}&\underline{0.0586}&0.0668 \\

\cellcolor{myblue}\model & \cellcolor{myblue}\textbf{0.0823} & \cellcolor{myblue}\textbf{0.1144} & \cellcolor{myblue}\textbf{0.1502} & \cellcolor{myblue}\textbf{0.0582} & \cellcolor{myblue}\textbf{0.0685} & \cellcolor{myblue}\textbf{0.0775} & \cellcolor{myblue}\textbf{0.0896} & \cellcolor{myblue}\textbf{0.1200} & \cellcolor{myblue}\textbf{0.1447} & \cellcolor{myblue}\textbf{0.0668} & \cellcolor{myblue}\textbf{0.0765} & \cellcolor{myblue}\textbf{0.0828} \\

\cellcolor{myred}improve
& \cellcolor{myred}39.49\%$^*$ & \cellcolor{myred}33.95\%$^*$ & \cellcolor{myred}27.29\%$^*$ & \cellcolor{myred}39.56\%$^*$ & \cellcolor{myred}37.83\%$^*$ & \cellcolor{myred}36.44\%$^*$ 
& \cellcolor{myred}16.06\%$^*$ & \cellcolor{myred}21.33\%$^*$ & \cellcolor{myred}7.66\%$^*$ & \cellcolor{myred}28.21\%$^*$ & \cellcolor{myred}30.55\%$^*$ & \cellcolor{myred}22.49\%$^*$ \\
  \bottomrule
  \toprule
  \multirow{2}{*}{Method} & \multicolumn{6}{c!{\vrule}}{Sports} & \multicolumn{6}{c}{Health} \\
   & RC@5 & RC@10 & RC@20 & NG@5 & NG@10 & NG@20 & RC@5 & RC@10 & RC@20 & NG@5 & NG@10 & NG@20 \\
  \midrule
  RNS& 0.0233 & 0.0368 & 0.0587 & 0.0153 & 0.0195 & 0.0250  &0.0360&0.0533&0.0777&0.0239&0.0295&0.0357\\
 DNS &0.0048& 0.0087& 0.0157 & 0.0027 & 0.0039 & 0.0056 &0.0243&0.0311&0.0388&0.0196&0.0217&0.0236  \\
 MixGCF & 0.0037 & 0.0059 & 0.0132 & 0.0023 & 0.0030 & 0.0048 &0.0207&0.0274&0.0344&0.0165&0.0186&0.0204 \\ 

   AdaSIR &0.0250&0.0354&0.0522&0.0168&0.0202&0.0245&0.0404&0.0603&0.0893&0.0275&0.0340&0.0413\\
  \hdashline
 GNNO  & \underline{0.0326} & 0.0480 & 0.0697 & 0.0206 & 0.0255 & 0.0310  &0.0458&0.0653&0.0955&0.0309&0.0372&0.0448\\

DNS+ & 0.0312 & 0.0486 & \underline{0.0713} & 0.0202 & 0.0258 & 0.0316 &0.0453&\underline{0.0681}&\underline{0.0968}&0.0316&0.0389&\underline{0.0461} \\
 MixGCF+   & 0.0323 & \underline{0.0489} & 0.0683 & \underline{0.0215} & \underline{0.0269} & \underline{0.0318}&\underline{0.0492}&0.0653&0.0950&\underline{0.0334}&0.0386&0.0461 \\
  SRNS &0.0315&0.0478&0.0680&0.0212&0.0265&0.0316&0.0469&0.0668&0.0922&0.0328&\underline{0.0393}&0.0456\\

\cellcolor{myblue}\model & \cellcolor{myblue}\textbf{0.0402} & \cellcolor{myblue}\textbf{0.0573} & \cellcolor{myblue}\textbf{0.0806} & \cellcolor{myblue}\textbf{0.0285} & \cellcolor{myblue}\textbf{0.0340} & \cellcolor{myblue}\textbf{0.0398} & \cellcolor{myblue}\textbf{0.0619} & \cellcolor{myblue}\textbf{0.0857} & \cellcolor{myblue}\textbf{0.1147} & \cellcolor{myblue}\textbf{0.0439} & \cellcolor{myblue}\textbf{0.0516} & \cellcolor{myblue}\textbf{0.0588} \\

\cellcolor{myred}improve & \cellcolor{myred}23.31\%$^*$ & \cellcolor{myred}17.18\%$^*$ & \cellcolor{myred}13.04\%$^*$ & \cellcolor{myred}32.46\%$^*$ & \cellcolor{myred}26.39\%$^*$ & \cellcolor{myred}25.18\%$^*$ 
& \cellcolor{myred}25.81\%$^*$ & \cellcolor{myred}25.84\%$^*$ & \cellcolor{myred}18.49\%$^*$ & \cellcolor{myred}31.44\%$^*$ & \cellcolor{myred}31.29\%$^*$ & \cellcolor{myred}27.54\%$^*$ \\

  \bottomrule
  \toprule
  \multirow{2}{*}{Method} & \multicolumn{6}{c!{\vrule}}{LastFM} & \multicolumn{6}{c}{KuaiRand} \\
   & RC@5 & RC@10 & RC@20 & NG@5 & NG@10 & NG@20 & RC@5 & RC@10 & RC@20 & NG@5 & NG@10 & NG@20 \\
  \midrule
  RNS& 0.0917 & 0.1483 & 0.2244 & 0.0591 & 0.0773 & 0.0966&0.0349 & 0.0606 & 0.111 & 0.0214 & 0.0298 & 0.0424
   \\
DNS &0.0782 &0.1209&0.1678&0.0515&0.0652&0.0769& 0.0349 & 0.0638 & 0.0976 & 0.0216 & 0.0310 & 0.0395
  \\
MixGCF&0.0735&0.1156&0.1675&0.0461&0.0597&0.0729& 0.0295 & 0.0413 & 0.0622 & 0.0216 & 0.0255 & 0.0307
 \\      
 AdaSIR&0.1061&0.1728&0.2552&0.0691&0.0905&0.1114& 0.0354 & 0.0606 & 0.1115 & 0.0218 & 0.0299 & 0.0428
\\
\hdashline
GNNO&\underline{0.1185}&\underline{0.1879}&\underline{0.2692}&\underline{0.0790}&\underline{0.1013}&\underline{0.1218} & \underline{0.0548} & 0.0847 & \underline{0.1426} & 0.0334 & 0.0447 & 0.0591
 \\
  
 DNS+& 0.1154 & 0.1768 & 0.2569 & 0.0745 & 0.0943 & 0.1145& 0.0542 & \underline{0.0923} & 0.1416 & \underline{0.0344} & \underline{0.0459} & \underline{0.0592}
  \\
 MixGCF+ & 0.1121 & 0.1682 & 0.2474 & 0.0694 & 0.0873 & 0.1073  & 0.0547 & 0.0847 & 0.1340 & 0.0326 & 0.0423 & 0.0547
 \\ 
    SRNS&0.1106&0.1739&0.2505&0.0728&0.0932&0.1126& 0.0477 & 0.0804 & 0.1239 & 0.0321 & 0.0425 & 0.0534
 \\
 
\cellcolor{myblue}\model  & \cellcolor{myblue}\textbf{0.1363} & \cellcolor{myblue}\textbf{0.2033} & \cellcolor{myblue}\textbf{0.2869} & \cellcolor{myblue}\textbf{0.0915} & \cellcolor{myblue}\textbf{0.1131} & \cellcolor{myblue}\textbf{0.1342} & \cellcolor{myblue}\textbf{0.0617} & \cellcolor{myblue}\textbf{0.0992} & \cellcolor{myblue}\textbf{0.1592} & \cellcolor{myblue}\textbf{0.0368} & \cellcolor{myblue}\textbf{0.0487} & \cellcolor{myblue}\textbf{0.0637} \\

\cellcolor{myred}improve & \cellcolor{myred}15.02\%$^*$ & \cellcolor{myred}8.20\%$^*$ & \cellcolor{myred}6.58\%$^*$ & \cellcolor{myred}15.82\%$^*$ & \cellcolor{myred}11.65\%$^*$ & \cellcolor{myred}10.18\%$^*$ 
& \cellcolor{myred}12.59\%$^*$ & \cellcolor{myred}7.48\%$^*$ & \cellcolor{myred}11.64\%$^*$ & \cellcolor{myred}6.98\%$^*$ & \cellcolor{myred}6.10\%$^*$ & \cellcolor{myred}7.60\%$^*$ \\

  \bottomrule
  \end{tabular}
  
  \label{tab:overall}
  \end{table*}

%% file: Tables_Algos/model_roub.tex
\begin{table*}[ht]
  \centering
  \caption{Performance of different sequential recommendation models on different real-world datasets using BCE loss~\cite{shannon1948mathematical}. The best scores are marked in bold fonts. RC and NG denote Recall and NDCG. * denotes p-value < 0.05 for paired t-tests.}
  \vspace{-1em}
  \setlength{\tabcolsep}{3.6pt}{
  \begin{tabular}{c!{\vrule}c!{\vrule}cccc!{\vrule}cccc!{\vrule}cccc}
  \toprule
  \multirow{2}{*}{Domain} & \multirow{2}{*}{Method} 
  & \multicolumn{4}{c!{\vrule}}{SASRec$_{BCE}$} 
  & \multicolumn{4}{c!{\vrule}}{LinRec$_{BCE}$} 
  & \multicolumn{4}{c}{FMLP4Rec$_{BCE}$} \\
   & & RC@5 & RC@10 & NG@5 & NG@10 & RC@5 & RC@10 & NG@5 & NG@10 & RC@5 & RC@10 & NG@5 & NG@10 \\
  \midrule
  \multirow{3}{*}{Beauty} 
  & RNS 
  & 0.0434 & 0.0612 & 0.0289 & 0.0347 & 0.0523 & 0.0751 & 0.0363 & 0.0435 & 0.0532 & 0.0800 & 0.0342 & 0.0429 \\
  & DNS+
  & 0.0550 & 0.0769 & 0.0355 & 0.0425 & 0.0617 & 0.0939 & 0.0420 & 0.0525 & 0.0635 & 0.0903 & 0.0432 & 0.0519 \\
  & \cellcolor{myblue}\model  
  & \cellcolor{myblue}\textbf{0.0823$^{*}$} & \cellcolor{myblue}\textbf{0.1144$^{*}$} & \cellcolor{myblue}\textbf{0.0582$^{*}$} & \cellcolor{myblue}\textbf{0.0685$^{*}$} 
  & \cellcolor{myblue}\textbf{0.0818$^{*}$} & \cellcolor{myblue}\textbf{0.1162$^{*}$} & \cellcolor{myblue}\textbf{0.0579$^{*}$} & \cellcolor{myblue}\textbf{0.0689$^{*}$} 
  & \cellcolor{myblue}\textbf{0.0872$^{*}$} & \cellcolor{myblue}\textbf{0.1198$^{*}$} & \cellcolor{myblue}\textbf{0.0621$^{*}$} & \cellcolor{myblue}\textbf{0.0726$^{*}$} \\ 
  \midrule
  \multirow{3}{*}{Toys} 
  & RNS & 0.0592 & 0.0865 & 0.0406 & 0.0496 & 0.0639 & 0.0860 & 0.0431 & 0.0505 & 0.0592 & 0.0870 & 0.0416 & 0.0505 \\
  & DNS+ & 0.0721 & 0.0963 & 0.0503 & 0.0581 & 0.0742 & 0.1050 & 0.0521 & 0.0622 & 0.0808 & 0.1056 & 0.0522 & 0.0600 \\
  
  & \cellcolor{myblue}\model  
  & \cellcolor{myblue}\textbf{0.0896$^{*}$} & \cellcolor{myblue}\textbf{0.1200 $^{*}$} & \cellcolor{myblue}\textbf{0.0668 $^{*}$} & \cellcolor{myblue}\textbf{0.0765$^{*}$} 
  & \cellcolor{myblue}\textbf{0.0963$^{*}$} & \cellcolor{myblue}\textbf{0.1241$^{*}$} & \cellcolor{myblue}\textbf{0.0688$^{*}$} & \cellcolor{myblue}\textbf{0.0777$^{*}$} 
  & \cellcolor{myblue}\textbf{0.0968$^{*}$} & \cellcolor{myblue}\textbf{0.1282$^{*}$} & \cellcolor{myblue}\textbf{0.0684$^{*}$} & \cellcolor{myblue}\textbf{0.0785$^{*}$} \\
  \midrule
  \multirow{3}{*}{Sports} 
  & RNS &0.0233 & 0.0368 & 0.0153 & 0.0195 &0.0261 & 0.0433 & 0.0159&0.0215&0.0261&0.0416&0.0163& 0.0213 \\
  &  DNS+ & 0.0312 & 0.0486 & 0.0202 & 0.0258 & 0.0323 & 0.0472 & 0.0220 & 0.0268 &0.0365&0.0494&0.0236&0.0278\\
  & \cellcolor{myblue}\model 
  & \cellcolor{myblue}\textbf{0.0402 }$^{*}$ & \cellcolor{myblue}\textbf{0.0573}$^{*}$ & \cellcolor{myblue}\textbf{0.0285}$^{*}$ & \cellcolor{myblue}\textbf{0.0340}$^{*}$
  & \cellcolor{myblue}\textbf{0.0475}$^{*}$ & \cellcolor{myblue}\textbf{0.0618}$^{*}$ & \cellcolor{myblue}\textbf{0.0313}$^{*}$ & \cellcolor{myblue}\textbf{0.0359}$^{*}$
  & \cellcolor{myblue}\textbf{0.0447}$^{*}$ & \cellcolor{myblue}\textbf{0.0635}$^{*}$ & \cellcolor{myblue}\textbf{0.0302}$^{*}$ & \cellcolor{myblue}\textbf{0.0363}$^{*}$ \\
  \bottomrule
  \end{tabular}}
  \label{tab:model_rob}
\end{table*}

%% file: Tables_Algos/loss_roub.tex
\begin{table*}[ht]
\centering
\caption{Performance of the other three sequential recommendation models on two real-world datasets using BPR loss~\cite{rendle2012bpr}. The best scores are marked in bold fonts. RC and NG denote Recall and NDCG. * denotes p-value < 0.05 for paired t-tests.}
\vspace{-1em}
\setlength{\tabcolsep}{3.6pt}{
\begin{tabular}{c!{\vrule}c!{\vrule}cccc!{\vrule}cccc!{\vrule}cccc}
\toprule
\multirow{2}{*}{Domain} & \multirow{2}{*}{Method} 
& \multicolumn{4}{c!{\vrule}}{SASRec$_{BPR}$} 
& \multicolumn{4}{c!{\vrule}}{GRU4Rec$_{BPR}$} 
& \multicolumn{4}{c}{Mamba4Rec$_{BPR}$} \\
 & & RC@5 & RC@10 & NG@5 & NG@10 & RC@5 & RC@10 & NG@5 & NG@10 & RC@5 & RC@10 & NG@5 & NG@10 \\
\midrule
\multirow{3}{*}{Beauty} 
& RNS 
 & 0.0416 & 0.0657 & 0.0300 & 0.0377 &0.0241&0.0398&0.0154&0.0204 & 0.0527 & 0.0791 & 0.0343 & 0.0428 \\
& DNS+
 & 0.0523 & 0.0760 & 0.0351 & 0.0427 &0.0340&0.0501&0.0231&0.0283& 0.0572 & 0.0760 & 0.0351 & 0.0427 \\
 & \cellcolor{myblue}\model 
  & \cellcolor{myblue}\textbf{0.0791}$^*$ & \cellcolor{myblue}\textbf{0.1109}$^*$ & \cellcolor{myblue}\textbf{0.0569}$^*$ & \cellcolor{myblue}\textbf{0.0671}$^*$ 
  & \cellcolor{myblue}\textbf{0.0612}$^*$ & \cellcolor{myblue}\textbf{0.0921}$^*$ & \cellcolor{myblue}\textbf{0.0425}$^*$ & \cellcolor{myblue}\textbf{0.0524}$^*$ 
  & \cellcolor{myblue}\textbf{0.0755}$^*$ & \cellcolor{myblue}\textbf{0.1122}$^*$ & \cellcolor{myblue}\textbf{0.0548}$^*$ & \cellcolor{myblue}\textbf{0.0667}$^*$ \\
\midrule
\multirow{3}{*}{Toys} 
& RNS & 0.0561 & 0.0865 & 0.0383 &0.0481  &0.0237&0.0371&0.0162&0.0205 & 0.0577 & 0.0860 & 0.0411 & 0.0501 \\
& DNS+ & 0.0633 & 0.0953 & 0.0445 & 0.0546 &0.0299&0.0469&0.0170&0.0225 & 0.0690 & 0.0947 & 0.0484 & 0.0568 \\
& \cellcolor{myblue}\model  
  & \cellcolor{myblue}\textbf{0.0911}$^*$ & \cellcolor{myblue}\textbf{0.1153}$^*$ & \cellcolor{myblue}\textbf{0.0658}$^*$ & \cellcolor{myblue}\textbf{0.0737}$^*$ 
  & \cellcolor{myblue}\textbf{0.0659}$^*$ & \cellcolor{myblue}\textbf{0.0901}$^*$ & \cellcolor{myblue}\textbf{0.0444}$^*$ & \cellcolor{myblue}\textbf{0.0522}$^*$ 
  & \cellcolor{myblue}\textbf{0.0932}$^*$ & \cellcolor{myblue}\textbf{0.1267}$^*$ & \cellcolor{myblue}\textbf{0.0656}$^*$ & \cellcolor{myblue}\textbf{0.0764}$^*$ \\

\midrule
\multirow{3}{*}{Sports} 
 &RNS &0.0270&0.0424&0.0171&0.0220&0.0129&0.0205&0.0082&0.0106&0.0278&0.0390&0.0197&0.0233\\
 &DNS+ &0.0284&0.0416&0.0204&0.0245&0.0171&0.0267&0.0106&0.0136&0.0320&0.0466&0.0215& 0.0262\\
  & \cellcolor{myblue}\model 
  & \cellcolor{myblue}\textbf{0.0433}$^*$ & \cellcolor{myblue}\textbf{0.0604}$^*$ & \cellcolor{myblue}\textbf{0.0293}$^*$ & \cellcolor{myblue}\textbf{0.0348}$^*$ 
  & \cellcolor{myblue}\textbf{0.0320}$^*$ & \cellcolor{myblue}\textbf{0.0469}$^*$ & \cellcolor{myblue}\textbf{0.0206}$^*$ & \cellcolor{myblue}\textbf{0.0253}$^*$ 
  & \cellcolor{myblue}\textbf{0.0441}$^*$ & \cellcolor{myblue}\textbf{0.0629}$^*$ & \cellcolor{myblue}\textbf{0.0308}$^*$ & \cellcolor{myblue}\textbf{0.0368}$^*$ \\
\bottomrule
\end{tabular}}
\label{tab:loss_rob}
\end{table*}

%% file: Tables_Algos/ablation.tex
\begin{table*}[ht]
\centering
\caption{Ablation study on three Amazon datasets. ($\downarrow$ \model) denotes the performance drop from the full \model \ method.}
\vspace{-1em}

\newcolumntype{Y}{>{\centering\arraybackslash}X}

\setlength{\tabcolsep}{1.5pt} 

\begin{tabularx}{\linewidth}{c!{\vrule}YYYY!{\vrule}YYYY!{\vrule}YYYY}
\toprule
\multirow{2}{*}{Variant}
& \multicolumn{4}{c!{\vrule}}{Beauty} 
& \multicolumn{4}{c!{\vrule}}{Toys} 
& \multicolumn{4}{c}{Sports} \\
& RC@5 & RC@10 & NG@5 & NG@10 & RC@5 & RC@10 & NG@5 & NG@10 & RC@5 & RC@10 & NG@5 & NG@10 \\
\midrule

\cellcolor{myblue}\model 
  & \cellcolor{myblue}\textbf{0.0823} & \cellcolor{myblue}\textbf{0.1144} & \cellcolor{myblue}\textbf{0.0582} & \cellcolor{myblue}\textbf{0.0685} 
  & \cellcolor{myblue}\textbf{0.0896} & \cellcolor{myblue}\textbf{0.1200} & \cellcolor{myblue}\textbf{0.0668} & \cellcolor{myblue}\textbf{0.0765} 
  & \cellcolor{myblue}\textbf{0.0402} & \cellcolor{myblue}\textbf{0.0573} & \cellcolor{myblue}\textbf{0.0285} & \cellcolor{myblue}\textbf{0.0340} \\
\hdashline

 w/o CD  &0.0697 & 0.0957 & 0.0491  & 0.0573 & 0.0860 & 0.1138 & 0.0600  & 0.0689 & 0.0368 & 0.0531& 0.0256  & 0.0308 \\
($\downarrow$\ \model) & 15.31\% & 15.64\% & 16.35\% & 16.35\% & 4.02\% & 5.17\% & 10.18\%  & 9.93\% & 8.46\% & 7.33\% & 10.18\% & 9.41\%
\\
\hdashline 

 w/o CD-DR & 0.0599& 0.0831 & 0.0434 & 0.0510 & 0.0819 & 0.1066 & 0.0585  & 0.0665 & 0.0343& 0.0503 & 0.0226  & 0.0278\\
($\downarrow$\ \model) & 27.22\% & 25.43\% & 27.36\% & 25.55\% & 8.59\% & 11.17\% & 12.43\%  & 13.07\% & 14.68\% & 12.22\% & 20.70\%  & 18.24\%
\\
\hdashline 
 
w/o CD-DR-MS&0.0550& 0.0769 & 0.0355  & 0.0425 & 0.0721& 0.0963  & 0.0503 & 0.0581 & 0.0312 & 0.0486 & 0.0202  & 0.0258
\\
($\downarrow$\ \model) & 33.17\% & 39.00\% & 32.78\% & 37.96\% & 19.53\% & 19.75\% & 24.70\% & 24.05\% & 22.39\% & 15.18\% & 29.12\%  & 24.12\%
\\
\bottomrule
\end{tabularx}
\label{tab:ablation}
\end{table*}

%% file: Tables_Algos/WTS.tex
\begin{table}[!t]
\centering
\caption{Performance of MDCNS under weak to strong setting.}
\label{tab:wts}
\vspace{-1em}
\newcolumntype{Y}{>{\centering\arraybackslash}X} 

\setlength{\tabcolsep}{3pt} 

\begin{tabularx}{\linewidth}{c!{\vrule}l!{\vrule}YYYY}
\toprule
\multirow{1}{*}{Domain} & \multirow{1}{*}{Method}  & RC@5 & RC@10 & NG@5 & NG@10 \\ 
\midrule

\multirow{4}{*}{Beauty} 
&RNS       & 0.0434 & 0.0612 & 0.0289 & 0.0347 \\
&DNS+      & 0.0550 & 0.0769 & 0.0355 & 0.0425 \\
& \cellcolor{myblue0}$\mathrm{\model_{wts}}$ & \cellcolor{myblue0}\underline{0.0760} & \cellcolor{myblue0}\underline{0.1140} & \cellcolor{myblue0}\underline{0.0542} & \cellcolor{myblue0}\underline{0.0663} \\
& \cellcolor{myblue}\model & \cellcolor{myblue}\textbf{0.0823} & \cellcolor{myblue}\textbf{0.1144} & \cellcolor{myblue}\textbf{0.0582} & \cellcolor{myblue}\textbf{0.0685} \\

\midrule

\multirow{4}{*}{Toys} 
&RNS       & 0.0592 & 0.0865 & 0.0406 & 0.0496 \\
&DNS+      & 0.0721 & 0.0963 & 0.0503 & 0.0581 \\
& \cellcolor{myblue0}$\mathrm{\model_{wts}}$ & \cellcolor{myblue0}\textbf{0.0942} & \cellcolor{myblue0}\underline{0.1195} & \cellcolor{myblue0}\underline{0.0650} & \cellcolor{myblue0}\underline{0.0731} \\
& \cellcolor{myblue}\model & \cellcolor{myblue}\underline{0.0896} & \cellcolor{myblue}\textbf{0.1200} & \cellcolor{myblue}\textbf{0.0668} & \cellcolor{myblue}\textbf{0.0765} \\

\midrule

\multirow{4}{*}{Sports} 
& RNS & 0.0233 & 0.0368 & 0.0153 & 0.0195 \\
& DNS+ & 0.0312 & 0.0486 & 0.0202 & 0.0258 \\
& \cellcolor{myblue0}$\mathrm{\model_{wts}}$ & \cellcolor{myblue0}\textbf{0.0433} & \cellcolor{myblue0}\textbf{0.0596} & \cellcolor{myblue0}\textbf{0.0299} & \cellcolor{myblue0}\textbf{0.0351} \\
& \cellcolor{myblue}\model & \cellcolor{myblue}\underline{0.0402} & \cellcolor{myblue}\underline{0.0573} & \cellcolor{myblue}\underline{0.0285} & \cellcolor{myblue}\underline{0.0340} \\

\midrule

\multirow{4}{*}{LastFM} 
&RNS& 0.0917 & 0.1483  & 0.0591 & 0.0773 \\
&DNS+& 0.1154 & 0.1768 & 0.0745 & 0.0943\\
& \cellcolor{myblue0}$\mathrm{\model_{wts}}$ & \cellcolor{myblue0}\textbf{0.1384} & \cellcolor{myblue0}\textbf{0.2078} & \cellcolor{myblue0}\textbf{0.0921} & \cellcolor{myblue0}\textbf{0.1144} \\
& \cellcolor{myblue}\model & \cellcolor{myblue}\underline{0.1363} & \cellcolor{myblue}\underline{0.2033} & \cellcolor{myblue}\underline{0.0915} & \cellcolor{myblue}\underline{0.1131} \\

\bottomrule
\end{tabularx}
\end{table}

%% file: sections/related_work.tex
\section{Related Works}
\label{sec:related work}
In this section, we present a literature review regarding sequential recommendation and negative sampling.
\subsection{Sequential Recommendation}
Sequential Recommendation (SR) aims to capture the dynamic evolution of user preferences to predict future engagements. 
The research landscape has witnessed a paradigm shift from early probabilistic transition models~\citep{rendle2010factorizing} and matrix factorization (MF) techniques~\cite{hidasi2016general} to sophisticated neural architectures. 
In the early deep learning era, GRU4Rec~\cite{hidasi2015session,hidasi2016parallel} pioneered the use of recurrent units, while convolutional architectures such as Caser~\cite{Tang2018PersonalizedTS} and NextItNet~\cite{yuan2019simple} were introduced to capture local temporal patterns. This focus subsequently shifted toward self-attention mechanisms, with SASRec~\cite{kang2018self} and Bert4Rec~\cite{sun2019bert4recsequentialrecommendationbidirectional} becoming the standards for modeling long-range dependencies. 
More recently, the frontier has been further extended by frequency-domain models like Filter-MLP~\cite{zhou2022filter}, as well as selective state-space architectures, exemplified by Mamba4Rec~\cite{liu2024mamba4rec} and RecMamba~\cite{yang2024uncovering}. 
Beyond architectural refinements, the community has increasingly embraced auxiliary learning paradigms. Notably, contrastive learning frameworks (e.g., CL4SRec and ICLRec~\cite{chen2022intent,qiu2022contrastive,xie2022contrastive}), causal inference~\cite{wang2022unbiased}, and distributionally robust optimization (DRO)~\cite{yang2024debiasing,zhou2023distributionally} have been explored as parallel paradigms to alleviate data sparsity and bolster model robustness in real-world recommendation scenarios.

\subsection{Negative Sampling} 
Negative sampling is fundamentally essential for sequential recommendation models trained on implicit feedback to obtain valid contrastive supervision~\cite{xu2022negative}. 
Early methodologies primarily depended on heuristic distributions, operationalized through uniform strategies like Random Negative Sampling (RNS)~\cite{rendle2012bpr} or frequency-biased approaches such as Popularity-based Negative Sampling (PNS)~\cite{chen2017sampling,zhou2020noise}. 
The paradigm subsequently shifted toward dynamic mechanisms. Notably, Dynamic Negative Sampling (DNS)~\cite{zhang2013optimizing} pioneered this transition by prioritizing unobserved items with elevated predicted scores. 
This trajectory was further enriched by adversarial generative architectures~\cite{wang2017irgan,park2019adversarial} and auxiliary-data-driven models, prominently featuring methods like SamWalker and MixGCF~\cite{ying2018graph, chen2019samwalker, ding2020simplify, huang2021mixgcf}. 
Currently, the community is actively addressing the intrinsic bottlenecks of hard negative mining from two primary perspectives. 
On one hand, frameworks including DNS+~\cite{shi2023theories} and GNNO~\cite{fan2023neighborhood} are dedicated to eliminating "false negative" noise. 
On the other hand, recent advances introduced by EXHANS and R$^2$NS~\cite{10.1145/3774904.3792360, 10.1145/3701551.3703535} reveal that the premature introduction of excessively hard negatives often triggers early-stage training collapse. 
Consequently, curriculum learning strategies~\cite{wang2021surveycurriculumlearning} have been increasingly adopted to dynamically pace the sampling difficulty throughout the training process.
However, existing methods still face three key issues: negative self-reinforcement loop, restricted sampling diversity and inefficient resource allocation.
Distinct from existing studies, we introduce a novel sampling method that disrupts the negative self-reinforcement loop via multi-source scoring, enhances sample diversity through divergence re-ranking, and capitalizes on computational resources via consensus distillation.

%% file: sections/conclusion.tex
\section{Conclusion}
\label{sec:conclusion}

In this paper, we presented MDCNS, a novel negative sampling framework inspired by ZPD theory that effectively addresses the negative self-reinforcement loop, restricted sampling diversity. and inefficient resource allocation. 
MDCNS integrates three key components: multi-source sampling, which breaks the vicious cycle of self-selection by leveraging peer and teacher signals; divergence re-ranking, which enhances sampling diversity by quantifying the prediction discrepancy between views; and consensus distillation, which optimizes training efficiency by aligning the model with the teacher's global consensus.
Extensive experiments across diverse datasets, backbones and loss functions demonstrate that MDCNS consistently outperforms state-of-the-art methods, achieving superior performance and generalization. 
While we fully utilized the computational cost, the multi-source architecture inevitably introduces additional overhead during the training phase. Future work will dedicate to exploring lighter and more efficient alternatives.


%% file: sections/acknowledgement.tex
\section*{Acknowledgements}
This work is supported in part by National Natural Science Foundation of China (No. 62422215 and No. 62472427), Major Innovation \& Planning Interdisciplinary Platform for the "DoubleFirst Class" Initiative, Renmin University of China, Public Computing Cloud, Renmin University of China, fund for building world-class universities (disciplines) of Renmin University of China, the Outstanding Innovative Talents Cultivation Funded Programs 2024 of Renmin University of China.